# Coupled Tank Non-linear System; Modeling and Level Control using PID and Fuzzy Logic Techniques

Akram Muntaser[1], Nagi Buaossa[2]
[1]College of Electronics Technology - Bani Walid muntasera1@udayton.edu
[2]University of dayton buaossan1@udayton.edu

**Abstract:** Liquid level control is very important in industrial field, where the liquid level is required, and to prevent overflows. The coupled-tank is a common system in industrial control processes. The system consists of two tanks connected together and the liquid flows between them. Tanks contain an inlet and outlet for each tank. The main principle of controlling this system is to maintain a constant level of liquid in both tanks when there are an inflow and outflow of liquid in each tank. To control liquid level in the coupled tank system, the mathematical model of the system had been derived and evaluated as a form of linear model. The mathematical model of coupled tank was developed to apply to both conventional and fuzzy control systems where the dynamic behavior of the system was considered. When the system had been designed the corresponding model was implemented in simulation by using *Matlab* and *Simulink* tools.

**Introduction**

The wide majority of classical control techniques has been developed for linear-time-invariant systems which are their behavior assumed to be almost known and well understood. However, many systems are mainly nonlinear and the dynamic behavior are complex and it cannot being known entirely. These types of model uncertainties are extremely difficult to control even with the conventional adaptive techniques.
Recently, new methods of control systems such predictive control and fuzzy logic control (FLC) have been implemented for process control [1], but few of them have succeeded for process technology due to lack of exact mathematical model and uncertain physical parameters. For this reason, fuzzy logic is a form of logic whose underlying modes of reasoning are approximate instead of exact reasoning [2, 3]. FLCs are controllers consisting of linguistic "*if-then*" rules that can be constructed using the knowledge of given parameters. Industries such as refinery, food making, and water treatment plants require liquid to be pumped, stored in tanks, and then pumped to other tanks. The control of liquid level and liquid flow between tanks is the main problem in these industries [4, 5]. The objective of the level controller is to maintain the liquid level at a certain value and be able to accept new set point values as desired. This paper presents a PID and FLC methods for level control of a coupled tank system. The advantage and disadvantages of both methods will be clearly stated when level control systems being subjected to the change of set points and liquid level disturbance.

**Coupled tank system structure**
The couple tank system consists of two tanks connected in series as shown in Fig. 1. The basic goal of controlling the coupled-tank system is to maintain the level of liquid in the tank constant when there is an inflow of liquid into the tank and/or outflow out of the tank. The input flow rate has to be regulated in order to keep the level at the required point. Here it's assumed that the liquid is non viscous and incompressible fluid.

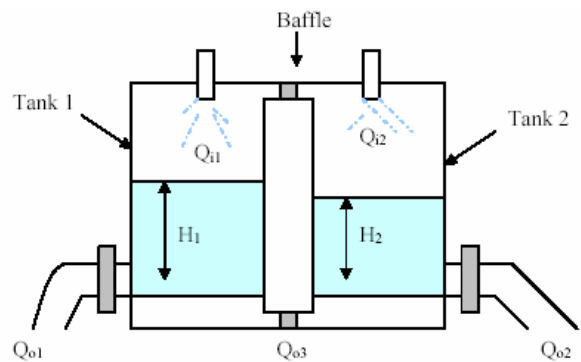

**Fig.1.** Structure of the coupled tank system

**Coupled tank system mathematical model**
The dynamic equation of coupled tank was derived based on Fig. 1. The model could be expressed as the second order with some linear approximation. Let H1 and H2 be the liquid level in each tank1 and tank2, respectively, which measured with respect to the corresponding outlet. If the input1 is the pump flow rate $Qi_1$, then the variable to be controlled would be the second state level H2. Tacking in consideration the disturbances caused by variations in the rate of flow out of the system caused by switching of the control valves. In view of a simple mass balance fundamentals, the change rate of fluid volume in each tank equals the net flow of fluid into the tank. Thus, for each of tank 1 and tank 2, the dynamic equation is developed as the following:

$$A_1 \frac{dH_1}{dt} = Q_{i1} - Q_{o1} - Q_{o3} \tag{1}$$

$$A_2 \frac{dH_2}{dt} = Q_{i2} - Q_{o2} - Q_{o3} \tag{2}$$

Where: $H_1, H_2$ = liquid level in tank 1 and tank 2, respectively
$A_1, A_2$ = cross-sectional areas of tank 1 and tank 2, respectively
$Qo3$ = liquid flow rate between tanks
$Q_{i1} Q_{i2}$ = pump flow rate into tank 1 and tank 2, respectively
$Q_{o1} Q_{o2}$ = flow rate of liquid out of tank 1 and tank 2, respectively

By using the Bernoulli equation for steady, non-viscous, incompressible liquid shows that the outlet flow in tank 2 is proportional to the square root of the head of water in this tank. Similarly, the flow between the two tanks is proportional to the square root of the differential head, where:

$$Q_{o1} = \alpha_1 \sqrt{H_1} \tag{3}$$

$$Q_{o2} = \alpha_2 \sqrt{H_2} \tag{4}$$

$$Q_{o3} = \alpha_3 \sqrt{H_1 - H_2} \tag{5}$$

Where $\alpha_1, \alpha_2$, and $\alpha_3$ are proportionality constant which depend on the coefficients of discharge, the cross-sectional area of each orifice and the gravitational constant. Substituting equations (3) (4) and (5) into equation (1) and (2), a set of nonlinear state equations which describe the system dynamics of the coupled tank apparatus are derived as:

$$A_1 \frac{dH1}{dt} = Q_{i1} - \alpha_1 \sqrt{H_1} - \alpha_3 \sqrt{H_1 - H_2} \tag{6}$$

$$A_2 \frac{dH2}{dt} = Q_{i2} - \alpha_2 \sqrt{H_2} - \alpha_3 \sqrt{H_1 - H_2} \tag{7}$$

**Linearized Perturbation Mode**
Suppose we set the inflows $Q_{i1}$ and $Q_{i2}$, the liquid level in the tanks is at some steady state levels $H_1$ and $H_2$. Consider small variations in each inflow, $q_1$ in $Q_{i1}$ and $q_2$ in $Q_{i2}$. Let the resulting perturbation in level be $h_1$ and $h_2$ respectively. From equations (6) and (7), the following equations can be derived:

For tank 1,
$$A_1 \frac{d(H_1 + h_1)}{dt} = (Qi_1 + q_1) - \alpha_1 \sqrt{H_1 + h_1} + \alpha_3 \sqrt{(H_1 - H_2 - h_1 - h_2)} \tag{8}$$

For tank 2,
$$A_2 \frac{d(H_2 + h_2)}{dt} = (Qi_2 + q_2) - \alpha_2 \sqrt{H_2 + h_2} + \alpha_3 \sqrt{(H_1 - H_2 - h_1 - h_2)} \tag{9}$$

Subtracting equations (6) and (7) from (8) and (9), the equation will become,

$$A_1 \frac{dh_1}{dt} = q_1 - \alpha_1 \left( \sqrt{H_1 + h_1} - \sqrt{H_1} \right) - \alpha_3 \left( \sqrt{(H_1 - H_2 - h_1 - h_2)} - \sqrt{H_1 - H_2} \right) \tag{10}$$

For small perturbation,

$$\sqrt{H_1 + h_1} = \sqrt{H_1} \left( \left( 1 + \frac{h_1}{H_1} \right) \right)^{0.5} \approx \sqrt{H_1} \left( \left( 1 + \frac{h_1}{2H_1} \right) \right)^{0.5} \tag{11}$$

Therefore consequently,

$$\sqrt{H_1 + h_1} - \sqrt{H_1} \approx \frac{h_1}{2\sqrt{H_1}} \tag{12}$$

Similarly,

$$\sqrt{H_2 + h_2} - \sqrt{H_2} \approx \frac{h_2}{2\sqrt{H_2}} \tag{13}$$

$$\left( \sqrt{H_1 - H_2 - h_1 - h_2} - \sqrt{H_1 - H_2} \right) \approx \frac{h_1 - h_2}{2\sqrt{H_1 - H_2}} \tag{14}$$

Abiding by this approximation, equations (15) and (16) are established,

$$A_1 \frac{dh_1}{dt} = q_1 - \frac{\alpha_1}{2\sqrt{H_1}} h_1 - \frac{\alpha_3}{\sqrt{H_1-H_2}}(h_1 - h_2) \tag{15}$$

$$A_2 \frac{dh_2}{dt} = q_2 - \frac{\alpha_2}{2\sqrt{H_2}} h_2 + \frac{\alpha_3}{2\sqrt{H_1-H_2}}(h_1 - h_2) \tag{16}$$

In equations (15) and (16), note that the coefficients of the perturbations in the level are functions of the steady state operating points $H_1$ and $H_2$. Note that the two equations can also be written in the form

$$A_1 \frac{dh_1}{dt} = q_1 - q_{o1} - \frac{\alpha_3}{\sqrt{H_1-H_2}}(h_1 - h_2) \tag{17}$$

$$A_2 \frac{dh_1}{dt} = q_2 - q_{o2} - \frac{\alpha_3}{\sqrt{H_1-H_2}}(h_1 - h_2) \tag{18}$$

Where $q_{o1}$ and $q_{o2}$ represent perturbation in the outflow at the drain pipes. This configuration is achieved by having the baffle raised at a small height. This allows flow of water from tank1 into tank2 and with this second order configuration, $h_2$ will be the process variable that is to be set which $q_1$ is the manipulated variable that is to be controlled. The other variable like $q_2$ will be assumed zero as the model is derived under the circumstances of no disturbance.

Performing Laplace transforms on equation (17) and (18) and assuming that initially all variables are at their steady state values,

$$A_1 s h_1(s) = q_1(s) - \left(\frac{\alpha_1}{2\sqrt{H_1}} + \frac{\alpha_3}{2\sqrt{H_1-H_2}}\right) h_1(s) + \frac{\alpha_3}{2\sqrt{H_1-H_2}} h_2(s) \tag{19}$$

$$A_2 s h_2(s) = q_2(s) - \left(\frac{\alpha_2}{2\sqrt{H_2}} + \frac{\alpha_3}{2\sqrt{H_1-H_2}}\right) h_2(s) + \frac{\alpha_3}{2\sqrt{H_1-H_2}} h_1(s) \tag{20}$$

By rearranging and rewriting in abbreviated manners,

$$(\tau_{1s} + 1) h_1(s) = k_1 q_1(s) + k_{12} h_2(s) \tag{19}$$

$$(\tau_{2s} + 1) h_2(s) = k_2 q_2(s) + k_{21} h_2(s) \tag{20}$$

Where

$$\tau_1 = \frac{A_1}{\frac{\alpha_1}{2\sqrt{H_1}} + \frac{\alpha_3}{2\sqrt{H_1-H_2}}} \qquad \tau_2 = \frac{A_2}{\frac{\alpha_2}{2\sqrt{H_2}} + \frac{\alpha_3}{2\sqrt{H_1-H_2}}} \tag{21}$$

$$K_1 = \frac{1}{\frac{\alpha_1}{2\sqrt{H_1}} + \frac{\alpha_3}{2\sqrt{H_1-H_2}}} \qquad K_2 = \frac{1}{\frac{\alpha_2}{2\sqrt{H_2}} + \frac{\alpha_3}{2\sqrt{H_1-H_2}}} \tag{22}$$

$$K_{12} = \frac{\frac{\alpha_3}{2\sqrt{H_1-H_2}}}{\frac{\alpha_1}{2\sqrt{H_1}} + \frac{\alpha_3}{2\sqrt{H_1-H_2}}} \qquad K_{21} = \frac{\frac{\alpha_3}{2\sqrt{H_1-H_2}}}{\frac{\alpha_2}{2\sqrt{H_2}} + \frac{\alpha_3}{2\sqrt{H_1-H_2}}} \tag{23}$$

Then, the transfer function of the linear second order dynamic system was derived as the following:

$$\frac{H_2(s)}{Q_1(s)} = \frac{K_1 K_2}{\tau_1 \tau_2 s^2 + (\tau_1 + \tau_2)s + (1 - K_{12} K_{21})} \tag{24}$$

The valve (pump actuator) can be also modelled as an important control element in the plant. The following differential equation describes the valve (pump actuator) dynamics as:

$$T_c \frac{dq(t)}{dt} + q_i(t) = Q_c(t) \tag{25}$$

Where,

$T_c$: is the time constant of the value/pump actuator
$q_i(t)$: is the time-varying input flow rate
$Q_c(t)$: is the computed or the commanded flow rate

Notice that, after the actuator is model merged with the plant, the commanded flow rate $Q_c$ from the actuator is actually the input flow rate $q_1$ into the first tank. Table. I. Shows the values of the system parameters obtained experimentally.

**Table I**: Coupled –Tank system parameter

| Name | Expression | Value | | |
|---|---|---|---|---|
| Cross Section Area Of the couple tank Reservoir | $A_1$ & $A_2$ | $32 cm^2$ | | |
| Proportionality Constant That Depends On Discharge Coefficient, Orifice Cross Sectional Area And Gravitational Constant | $\alpha_1$ Subscript denotes Which tank it refers | $\alpha_1$ | $\alpha_2$ | $\alpha_3$ |
| | | 14.30 $cm^{3/2}/sec$ | 14.30 $cm^{3/2}/sec$ | 20.00 $cm^{3/2}/sec$ |
| Pump motor (valve) Time constant | $T_c$ | 1sec(can be adjusted ) | | |
| Maximum allowable Volumetric flow rate Into the plant | $Qi_{max}$ | $300\ cm^3/s$ | | |

If we consider small variations of flow $q_1$ and the level $h_2$ at the steady state flow $Q_{i1}$ and the level $H_2$, then the time constants $\tau_1$, $\tau_2$ and couple tank parameters could be substituted as the following:

$$\tau_1 = 7.445 \qquad \tau_2 = 6.2$$
$$K_1 = 0.23267 \qquad K_2 = 0.1939$$
$$K_{12} = 0.6453 \qquad K_{21} = 0.5378$$

Then, the final transfer function (9) of the couple tanks can be rewritten in numerical values as:

$$\frac{H_2(s)}{Q_1(s)} = \frac{0.045}{46.159s^2 + 13.645s + 0.65295} \qquad (26)$$

**Controller design**

First, let us start with PID controller to compare at the end with the fuzzy logic controller. In order to find a better performance of control system, PID controller must be designed to achieve small overshoot, zero steady state error and fast response. In this work, PID controller parameters could be tuned using closed-loop Ziegler-Nichols method [6,7], where parameters obtained as follows:

**Table II.** PID controller parameters using the Z-N method

| Controller type | Proportional Gain $K_p$ | Integral Time $T_i$ | Derivative time Td |
|---|---|---|---|
| Proportional + Integral +derivative | $K_p = K_u/1.7 =$ $142/1.7 = 83.5$ | $T_i = T_u/2 =$ $11.5/2 = 5.75$ $K_i = K_p/T_i = 14.5$ | $T_d = T_u/8 =$ $11.5/8 = 1.437$ $K_d = K_p . T_d = 120$ |

Fig.2. shows the Simulink block diagram for the process and PID controller design.

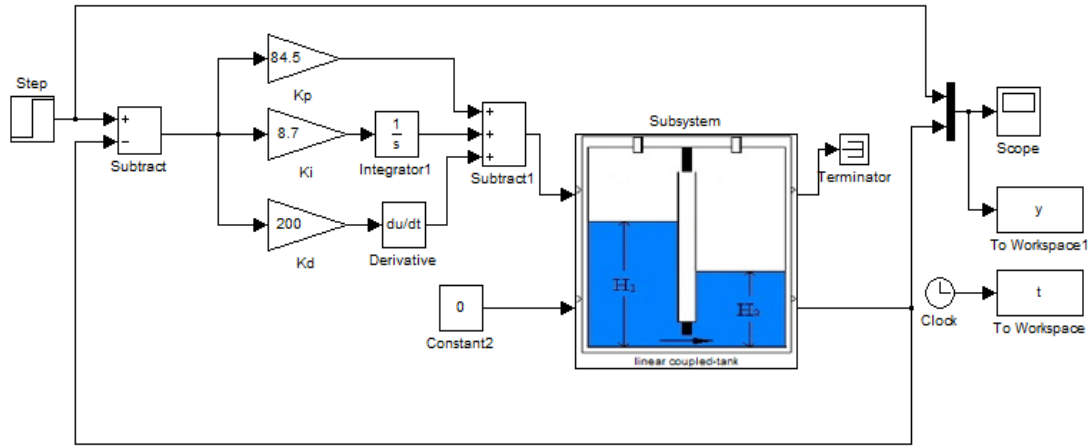

**Fig.2.** Schematic of the system with the PID controller

Next, Fuzzy controller should be designated and compared with classical PID control. The model of the system along with the Fuzzy Controller using MATLAB is shown in Fig. 3.

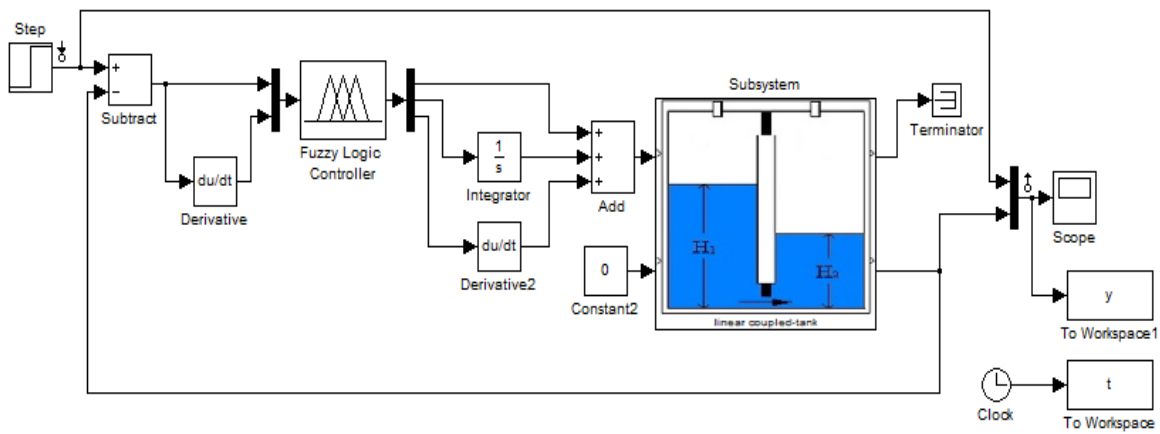

**Fig.3.** Schematic of the system with the PID controller

The number of necessary fuzzy sets and their limits were designed based upon the experience obtained in the process. The standard fuzzy set consists of three parts: fuzzification, *Inference* engine and defuzzification [8].
**Fuzzification** converts a crisp number into fuzzy sets within a universe of discourse. The triangular membership functions with five linguistic values with two inputs such as error (E) and change of error (CE) is used as shown in Figs. 4 and 5.

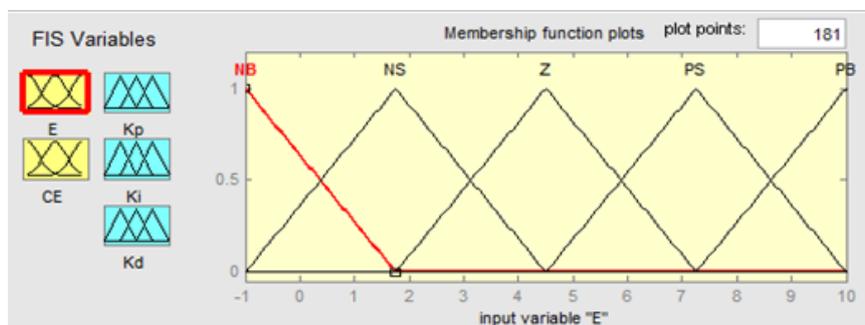

**Fig.4.** Membership function of the input variable E(t) of the PIDFLC

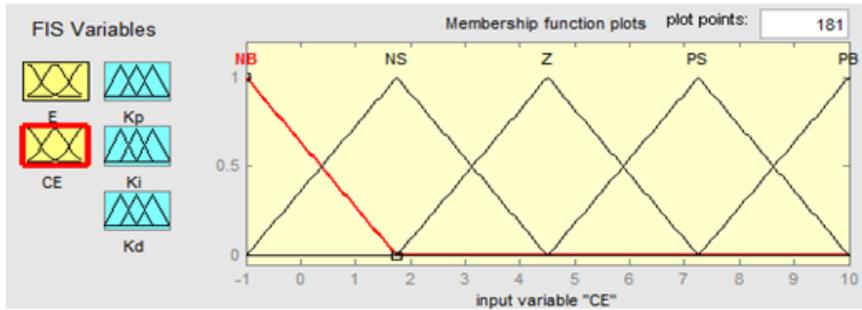
**Fig.5.** Membership function of the input variable CE(t) of the PIDFLC

The output of fuzzy sets correspond to the parameters of the classical PID controller where the output *Kp*, *Ki*, and *Kd*, also are the triangular shape which selected to present the range of the parameters as shown in Figures 6, 7 and 8, respectively.

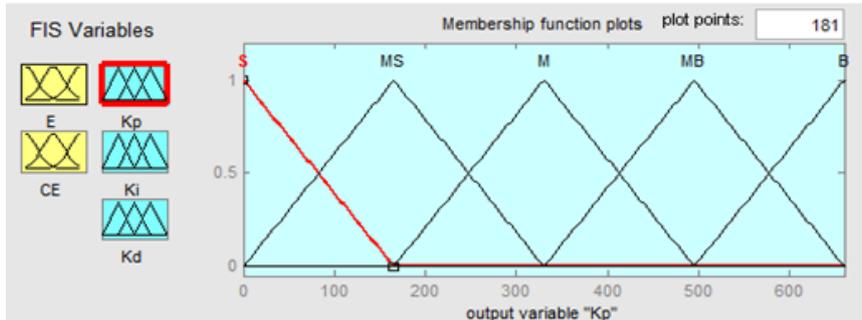
**Fig.6.** Membership function of the output variable (*Kp*) of the PIDFLC

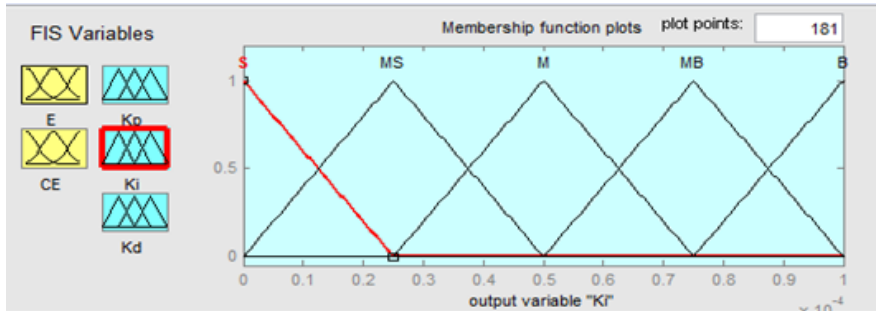
**Fig.7.** Membership function of the output variable (*Ki*) of the PIDFLC

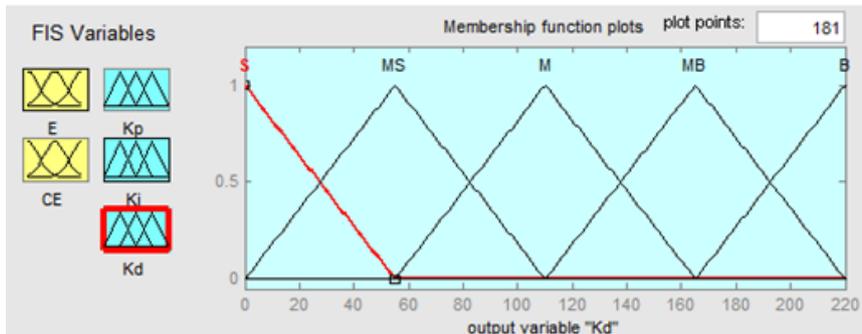
**Fig.8.** Membership function of the outputs variable (*Kd*) of the PIDFLC

**Inference engine:** It is the core of the fuzzy control and is constructed from expert knowledge and experience of the human. Based on the knowledge gained by analyzing the feedback control system inference engine is given in Table 2 where 25 rules are used. The fuzzy logic control rules will be as the following:

Table III. Rules of the fuzzy inference engine

| E/CE | NB | NS | Z  | PS | PB |
|------|----|----|----|----|----|
| NB   | S  | S  | S  | MS | M  |
| NS   | S  | S  | MS | M  | MB |
| Z    | S  | MS | M  | MB | B  |
| PS   | MS | M  | MB | B  | B  |
| PB   | M  | MB | B  | B  | B  |

The input linguistic values are Negative Big (NB), Negative Big (NS), Zero (Z), Positive Small (PS), and positive big (PB). The output values are small (S), medium (M), big (B), medium small (MS), and medium big (MB).
**Defuzzification**: It converts the fuzzy value into crisp value. In this study, center gravity method is used [9,10]. The triangular membership function with five linguistic values have been verified.

**Experimental results**
The model was tested on environmental of simulink/matlab whereas PID and FLC have been simulated under some circumstances such as step and disturbance input signals. The following results show the comparison between the classical PID and fuzzy PID controller. Fig.9. illustrates the comparison between PID classic and PID fuzzy logic controller performance. Note that the graph with red line represents the system response under PID fuzzy controller, while blue dotted line represents the system response under the PID controller.

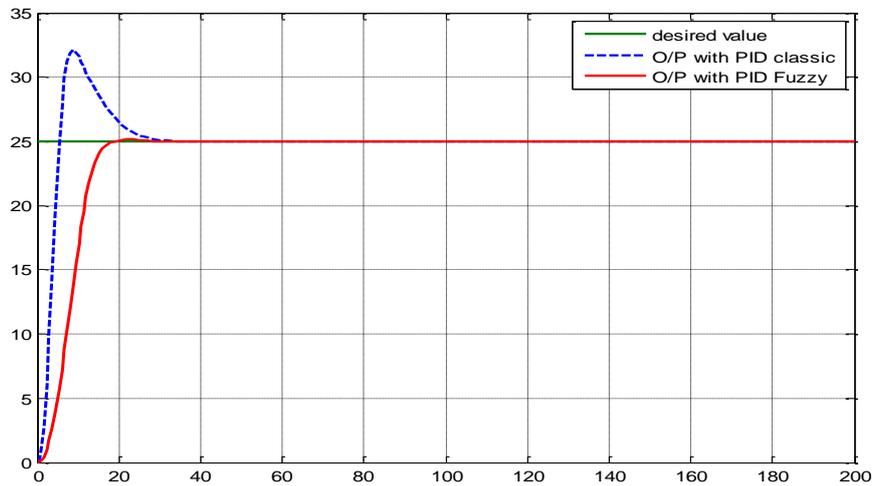

**Fig.9.** System response under PID and PIDFL controllers

**Disturbances (load) response:**
If we applied amount of disturbance for both PID and fuzzy PID, we would obtain unexpected results. Fig.10. shows the the system response under PID controller returned to the reference point after a short period after the disturbance happened to the system. Whereas the fuzzy controller unable to return to the reference value after the disturbance.

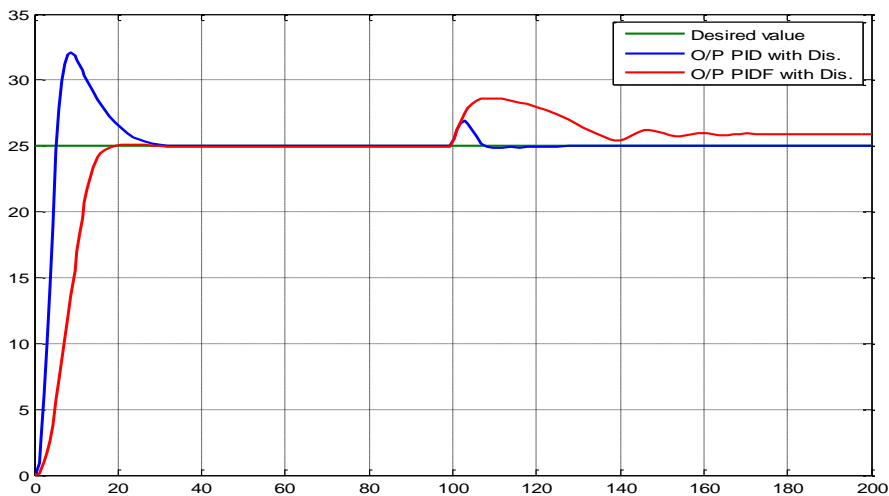

**Fig.10.** Disturbance effect on the PID and PIDFL controllers

The comparison of the control results from these two systems indicated that the fuzzy logic controller significantly reduced overshoot and steady state error (which means, in this system, the overflow and level stability, respectively). The comparison of the PID and FLC results are shown on table III below. The overall performance may be summarized as:

**Table III.** Comparison results of PID and FLC

| Controller | Overshoot | Rise time | Settling time | S-S Error |
|---|---|---|---|---|
| PID controller | 28% | 3.72 | 24.9 [Sec] | 0 |
| PID fuzzy logic controller | 0% | 9.95 | 16 [Sec] | 0 |

**Conclusion**

The basic goal of controlling the coupled-tank system is to maintain the level of liquid in the tank constant when there is an inflow of liquid into the tank and/or outflow out of the tank. The input flow rate has to be regulated in order to keep the level at the required point. Here it's assumed that the liquid is non viscous and incompressible fluid.

Unlike the conventional PID controller the Fuzzy Logic Controller has better performance on the system response. Adjusted FLC using a small number of rules and straightforward implementation has been proposed to solve a problem of level control with uncertain parameters commonly found in real industries. Moreover, the FLC can be easily programmed into many currently available industrial process operations. The FLC on a level control problem with promising results can be applied to an entirely different industrial level controlling apparatus. However, the main drawback of the fuzzy controller was unable to eliminate the effects of disturbance, and may become non robust controller under these circumstances. As a future work one can develop an FLC for a couple tanks system as adaptive Fuzzy Logic Controller, which can give high performance for control systems and overcome the problem of disturbances.

**References**

[1] J. Jin, H. Huang. "Study on fuzzy self-adaptive PID control system of biomass boiler drum water ". Journal of bio-energy systems, 2013.

[2] G, Himanshu, and OP Verma. "Intelligent controller for coupled tank system." International Journal of Advanced Research in Computer Science and Software Engineering 2.4 (2012).

[3] A. Muntaser, H. Elwarfalli, J. Kumar and G. Subramanyam, "Development of advanced energy storage system using fuzzy control," 2016 IEEE National Aerospace and Electronics Conference (NAECON) and Ohio Innovation Summit (OIS), 2016, pp. 179-182, doi: 10.1109/NAECON.2016.7856795.

[4] L. A. Zadeh. "Fuzzy sets: Inform. &Cont. ". Vol. 8, pp. 338-353, 1965.

[5] A. Muntaser, H. Elwarfalli, A. Suleiman and G. Subramanyam, "Design and implementation of conventional (PID) and modern (Fuzzy logic) controllers for an energy storage system of hybrid electric vehicles," 2017 IEEE National Aerospace and Electronics Conference (NAECON), 2017, pp. 267-270, doi: 10.1109/NAECON.2017.8268783.

[6] H. Elwarfalli, A. Muntaser, J. Kumar and G. Subramanyam, "Design and implementation of PI controller for the hybrid energy system," 2016 IEEE National Aerospace and Electronics Conference (NAECON) and Ohio Innovation Summit (OIS), 2016, pp. 170-172, doi: 10.1109/NAECON.2016.7856793.

[7] Elwarfalli, H., Abdalla Suleiman, M. M., & Muntaser, A. (2019, March). SCADA Control Levels for Two Tanks Process. In International Conference on Technical Sciences (ICST2019) (Vol. 6, p. 04).

[8] J. Hines. "MATLAB supplement to fuzzy and neural approaches in engineering ". John Wiley & Sons Inc., 1997.

[9] Muntaser, A., Suleiman, A. A., & Lesewed, A. A. "Synthesis of Fuzzy Control System for Coupled Level Tanks". 2015 International Conference on Information Science and Management Engineering (ICISME 2015) ISBN: 978-1-60595-303-8

[10] J. Jin, H. Huang. "Study on fuzzy self-adaptive PID control system of biomass boiler drum water ". Journal of bio-energy systems, 2013.